\documentclass[%
 reprint,
 amsmath,amssymb,
 aps,
floatfix,
]{revtex4-2}
\usepackage{graphicx}% Include figure files
\usepackage{dcolumn}% Align table columns on decimal point
\usepackage{bm}
\usepackage{xcolor}
\usepackage{soul}% bold math
\begin{document}
\preprint{APS/123-QED}

\title{Demonstration of Frequency Stability \\limited by Thermal Fluctuation Noise in Silicon Nitride Nanomechanical Resonators}% Force line breaks with \\

\author{Chang Zhang, Raphael St-Gelais}
\altaffiliation{raphael.stgelais@uottawa.ca}%Lines break automatically or can be forced with \\
\affiliation{
Department of Mechanical Engineering, University of Ottawa, Ottawa, ON, Canada\\
$^*$raphael.stgelais@uottawa.ca
}

\begin{abstract}
%The frequency stability of nanomechanical resonators (NMR) dictates the fundamental performance limit of sensors that relate physical perturbations to a resonance frequency shift. While the contribution of thermomechanical noise to frequency stability was understood recently, thermal fluctuation noise has attracted less attention despite being the ultimate performance limit of temperature sensing. We provide a model for the frequency stability of NMR considering both additive phase noise (i.e., thermomechanical and detection noises) and thermal fluctuation noise. We then experimentally demonstrate optimized NMR achieving frequency stability limited by thermal fluctuation noise. Our work shows that current models for NMR frequency stability can be incomplete. It also paves a way for NMR radiation detectors to reach the unattained fundamental detectivity limit of thermal-based radiation sensing.

The frequency stability of nanomechanical resonators (NMR) dictates the performance level of many state-of-the-art sensors (e.g., mass, force, temperature, radiation) that relate an external physical perturbation to a resonance frequency shift. While this is obviously of fundamental importance, accurate models and understandings of sources of frequency instability are not always available. The contribution of thermomechanical noise to frequency stability has been well studied in recent years and is often the fundamental performance limitation. Frequency stability limited by thermal fluctuation noise has attracted less interest but is nevertheless of fundamental importance notably in temperature sensing applications. In particular, temperature-sensitive NMR have become promising candidates for replacing traditional bolometers in infrared radiation sensing. However, reaching the ultimate detectivity limit of thermal radiation sensors requires their noise to be dominated by fundamental thermal fluctuation, which has not been demonstrated to date. In this work, we first develop a theoretical model for computing the frequency stability of NMR by considering the effect of both additive phase noise (i.e., thermomechanical, and experimental detection noise) and thermal fluctuation noise in a close-loop frequency tracking scheme. We thereafter validate this model experimentally and observe thermal fluctuation noise in SiN drum resonators of various sizes. Our work shows that by using resonators of specific characteristics---such as high temperature sensitivity, high mechanical quality factors, and high mass-to-thermal-conductance ratio---one can minimize additive phase noise below thermal fluctuation noise. This paves the way for NMR-based radiation sensors that can reach the fundamental detectivity limit of thermal radiation sensing and outperform existing technologies.

\end{abstract}

\maketitle

%\tableofcontents
Nanomechanical resonators (NMR) frequency stability is the fundamental quantity dictating the performance of sensors measuring physical signals through resonance frequency shifts (e.g., mass \cite{mass1,mass2,mass3,mass4}, force \cite{force1,force2}, and thermal \cite{thermal1,graphene,thermal3,zhang_fast_2019,laurent_12textensuremath-ensuremathmumathrmm-pitch_2018,zhang_radiative_2020} sensors). Hence, identifying the source of noise that dominates frequency fluctuations in the absence of a signal is fundamental for understanding the performance limit of NMR. Early theoretical investigation of various sources of noise in NMR was proposed by Vig $et~al.$ \cite{Vig1999} and Cleland $et~al.$ \cite{Cleland2002}. These theoretical works provided a comprehensive picture of multiple sources of noise in NMR, including surface diffusion, absorption-desorption, thermomechanical, detection, and thermal fluctuation noises. Thermomechanical noise was later identified as a dominant source of noise in many cases. Substantial efforts were therefore devoted to resolving the frequency stability limit imposed by thermomechanical noise  \cite{sansa_frequency_2016,roy_improving_2018,thermo2,thermo1}, and to predicting its effect after processing by various frequency tracking schemes. Demir \cite{demir_understanding_2021} recently provided a theoretical model for predicting the frequency stability of resonators, which considers the combined effect of thermomechanical and detection noise in  commonly employed phase-locked loop (PLL) frequency tracking. Such theoretical model was later validated experimentally by Sadeghi $et~al.$ \cite{sadeghi_frequency_2020}, confirming that thermomechanical and detection noises can be the dominant noises in high-stress silicon nitride string resonators.

\begin{figure}[!htb]
\includegraphics[scale=1]{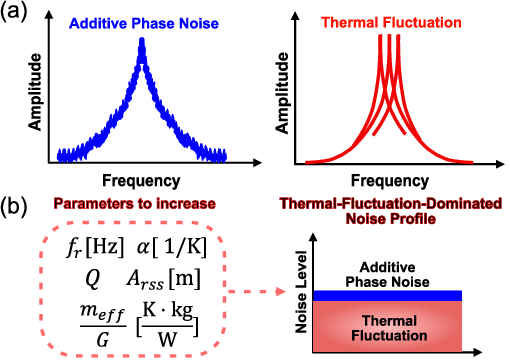}
\caption{\label{fig:one} \textbf{Additive phase noise vs. thermal fluctuation} (a) Illustration of the characteristic difference between additive phase noise and thermal fluctuation. (b) Increase in temperature coefficient $\alpha$, resonance frequency $f_r$, mechanical Q-factor, vibration amplitude $A_{rss}$, mass-to-thermal-conductance ratio $m_{eff}/G$ of the nanomechanical resonator (NMR) minimizes additive phase noise relative to thermal fluctuation noise.}
\end{figure}

While thermomechanical noise was proven to dominate frequency fluctuation in many high-Q factor NMR, achieving regimes in which thermal fluctuation noise would dominate remains of high interest, especially in temperature sensing applications. In temperature sensors, the smallest temperature that theoretically can be measured is ultimately dictated by thermal fluctuation noise. This limit is of particular interest in the field of radiation detection, as it dictates the fundamental detectivity limit that can ultimately be reached by thermal-based radiation detectors \cite{Rogalski2003,snell_heat_2022}, or bolometers. In mechanical resonator-based radiation sensors \cite{thermal1,graphene,thermal3,zhang_fast_2019,laurent_12textensuremath-ensuremathmumathrmm-pitch_2018,zhang_radiative_2020}, thermal fluctuation is typically not the dominant source of noise, which means that their ultimate limit of performance has not been reached. Likewise, the fundamental limit has not been reached in traditional (i.e., electrical) bolometers since Johnson-Nyquist noise always dominates over thermal fluctuation noise \cite{Rogalski2003}. NMR are potentially interesting due to their immunity to Johnson-Nyquist noise, but so far were found to be limited instead by thermomechanical, detection \cite{thermal1,thermal3,zhang_fast_2019,laurent_12textensuremath-ensuremathmumathrmm-pitch_2018,zhang_radiative_2020}, or flicker noise \cite{graphene}.

In the current work, we experimentally demonstrate low-stress, high Q-factor SiN drum resonators in which frequency instability is minimized down to fundamental thermal fluctuation noise. We also include thermal fluctuation noise with recently proposed models for frequency stability within a closed-loop frequency tracking scheme \cite{demir_understanding_2021}. Our results and model, therefore, pave the way for radiation sensors that could reach the fundamental detectivity limit of physical radiation detectors. 

We separate the sources of frequency fluctuation in a drum resonator into two categories, namely, the additive phase noise $S_{y,add}(\omega)$ which includes thermomechanical $S_{y,mech}(\omega)$ and detection noise $S_{y,det}(\omega)$; and thermal fluctuation noise $S_{y,th}(\omega)$. Here $y$ represents fractional frequency $\langle \delta f/f_r \rangle$ fluctuations. We also define the intrinsic noise spectral density $S_{y}^{int}(\omega)$ before it is modified by the experimental frequency measurement scheme (e.g., open-loop, phase-lock loop frequency tracking, or self-sustained oscillation). For a given eigen frequency $f_r$, the intrinsic frequency fluctuation caused by additive phase noise $S_{y,add}^{int}(\omega)$ is therefore \cite{demir_understanding_2021}:   
\begin{equation}
\begin{aligned}
S_{y,add}^{int}(\omega) &= S_{y,mech}^{int}(\omega)+S_{y,det}(\omega) \\
& =\frac{k_BT}{8\pi^3m_{eff}f_r^3QA_{rss}^2}|H_{mech}(j\omega)|^2\\
&+\frac{k_BT}{8\pi^3m_{eff}f_r^3QA_{rss}^2}\kappa_d^2\\
& = \frac{k_BT}{8\pi^3m_{eff}f_r^3QA_{rss}^2}(|H_{mech}(j\omega)|^2+\kappa_d^2),\\
\label{eq:1}
\end{aligned}
\end{equation}
where $H_{mech}(j\omega)=1/(1+j\omega \tau_{mech})$ is a one-pole low pass filter accounting for the mechanical time constant $\tau_{mech}=Q/(\pi f_r)$ of the resonator, $k_B$ is the Boltzmann constant, $T$ is the eigenmode temperature, $m_{eff}$ is the mode effective mass, $Q$ is the mechanical quality factor, $A_{rss}$ is the vibration amplitude in steady-state and $\kappa_d$ is a dimensionless parameter scaling the level of detection noise relative to thermomechanical noise, as described in \cite{demir_understanding_2021}. More specifically, thermomechanical noise is resolved above detection noise when $\kappa_d<1$. 

In turn, the intrinsic frequency fluctuation caused by thermal fluctuation \cite{Vig1999,snell_heat_2022} is:
\begin{equation}
S_{y,th}^{int}(\omega) = \frac{2k_BT^2\alpha^2}{\pi G}|H_{th}(j\omega)|^2,
\label{eq:2}
\end{equation}
where $G$ is the total thermal conductance \cite{zhang_radiative_2020} in W/K between the resonator and its environment, $H_{th}(j\omega)=1/(1+j\omega \tau_{th})$ is a one-pole filter accounting for the thermal response time $\tau_{th}=C_{th}/G$ of the resonator, and $C_{th}$ is the heat capacity of the resonator in J/K. An analytical model for $G$ (and hence $\tau_{th}$) in drum resonators is developed in \cite{zhang_radiative_2020} and is used throughout this work. $\alpha$ is the temperature coefficient of fractional frequency shifts, in $\mathrm{K^{-1}}$. For a drum resonator, a reasonable approximation ($<20$\% error relative to finite element modeling \cite{zhang_radiative_2020}) of this parameter is:
\begin{equation}
\alpha \approx \frac{E\alpha_T}{2\sigma(1-\nu)},
\label{eq:3.1}
\end{equation}
in which $E$ is Young's modulus, $\alpha_T$ is the drum material thermal expansion coefficient, $\sigma$ is the built-in tensile stress, and $\nu$ is the Poisson ratio of the drum resonator material. In Eq.~(\ref{eq:2}), $T$ denotes the drum resonator material temperature, whereas $T$ in Eq.~(\ref{eq:1}) denotes the temperature of the eigenmode of frequency $f_r$. We use the symbol $T$ interchangeably in this work since both are assumed to be $\approx300$ K.

A key difference between the additive phase noise and thermal fluctuation noise is schematized in Fig.~\ref{fig:one}(a), in which the additive phase noise manifests itself as amplitude fluctuation, which then contributes to the frequency noise via Robins' formula \cite{robin}. Conversely, thermal fluctuations make the resonance peaks of the NMR fluctuate directly in the frequency domain by affecting the stiffness of the drum material. We note that one can change the level of additive phase noise by utilizing different eigenmodes of the resonators [see Eq.~(\ref{eq:1})] since $f_r$, $Q$ are mode-dependent. Likewise, additive phase noise can be minimized by increasing the vibration amplitude $A_{rss}$ (within the linear actuation regime). On the contrary, mode and amplitude changes do not affect the level of thermal fluctuation [see Eq.~(\ref{eq:2})], which depends primarily on drum resonator geometric, material and heat transfer properties (e.g., $\alpha$ and $G$).

To compare the relative contributions of $S_{y,mech}^{int}(\omega)$ and $S_{y,th}^{int}(\omega)$ with respect to the overall frequency fluctuation, we define a dimensionless ratio $\gamma$:
\begin{equation}
\begin{aligned}
\gamma &= \frac{S_{y,th}^{int}(0)}{S_{y,mech}^{int}(0)} \\
& = 16\pi^2TQ\frac{m_{eff}}{G}\cdot\alpha^2A_{rss}^2f_r^3\\
\label{eq:3}
\end{aligned}
\end{equation}
This ratio ignores, for simplicity, the effect of both thermal $H_{th}(j\omega)$ and mechanical $H_{mech}(j\omega)$ response filters by setting $\omega=0$. A value of $\gamma\approx1$ indicates thermal fluctuation and additive phase noise are at a similar level. If $\gamma$ is significantly larger than 1 and detection noise is minimized (i.e., $\kappa_d<<1$), a thermal-fluctuation-dominated noise profile can be achieved. 

\begin{figure}[!htb]
\includegraphics[scale=1.13]{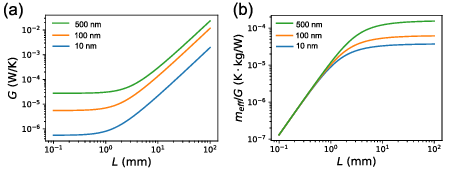}
\caption{\label{fig:five} \textbf{Effect of dimensions on heat transfer quantities in silicon nitride drum resonators,} namely (a) $G$ and (b) $\frac{m_{eff}}{G}$.}
\end{figure}

By examining $\gamma$, we find that thermal-fluctuation-dominated noise profile can be achieved by maximizing the values of $Q$, $\frac{m_{eff}}{G}$, $\alpha$, $A_{rss}$ and $f_r$. Among these listed parameters, the mass-to-thermal-conductance ratio $\frac{m_{eff}}{G}$ is likely the most unintuitive. Hence, in Fig.~\ref{fig:five}, we illustrate this ratio as a function of the dimensions of a square SiN drum resonators of side length $L$ and thickness $t$. Values of $G$ are presented in Fig.~\ref{fig:five}(a) and are calculated using the model developed in \cite{zhang_radiative_2020} which includes conductive and radiative heat transfer. We note that for relatively small drum resonators ($L<1$ mm) where heat transfer occurs mostly by conduction, both $G$ and $m_{eff}$ scale with the thickness, such that $t$ cancels out in the $\frac{m_{eff}}{G}$ ratio [see Fig.~\ref{fig:five}(b)]. Conversely, as $L>1$ mm, radiative heat transfer dominates, and $L$ now cancels, making the curves plateau in Fig.~\ref{fig:five}(b). 

In summary, considering this trend in $\frac{m_{eff}}{G}$ and the expression of $\gamma$, we find that minimizing additive phase noise below thermal fluctuation requires (i) a temperature sensitive resonator (i.e., high $\alpha$) which can be maximized using a low-stress material as suggested in Eq.~(\ref{eq:3.1}), (ii) a large drum resonator (i.e., large $L$) to maximize $\frac{m_{eff}}{G}$, (iii) a high order (i.e., high $f_r$) and high Q eigenmode, and (iv) excitation at high amplitude $A_{rss}$ within the limits of linear actuation. We finally note that, despite the trend observed in Fig.~\ref{fig:five}(b), increasing $t$ beyond $\approx100~\mathrm{nm}$ to maximize $\frac{m_{eff}}{G}$ is likely not a practical approach since high $t$ can also be detrimental to Q-factor \cite{schmid_Q,Q-fa}.   

To evaluate the spectral density of the frequency fluctuations $S_{y}(\omega)$ in a practical experimental setting, we must also consider the effect of the measurement scheme. A phase-locked loop (PLL) frequency tracking scheme, such as in the current work, includes a proportional-integral controller with proportional gain $K_p$ and integral gain $K_i$, and an input demodulator filter of time constant $\tau_{demod}$. As shown in \cite{demir_understanding_2021}, the PLL frequency tracking scheme imposes filters on the additive phase noises of the resonators (i.e., the thermomechanical and detection noises):
\begin{equation}
H_{mech}^{PLL}(j\omega) = \frac{(j\omega K_p + K_i)H_L(j\omega)}{-\omega^2+\frac{j\omega}{\tau_{mech}}+(j\omega K_p + K_i)H_L(j\omega)},
\label{eq:4}
\end{equation}
\begin{equation}
H_{det}^{PLL}(j\omega) = \frac{H_{mech}^{PLL}(j\omega)}{H_{mech}(j\omega)},
\label{eq:5}
\end{equation}
where $H_L(j\omega) = 1/(1+j\omega \tau_{demod})$ is the demodulator filter. Since $S_{y,mech}(\omega)$ and $S_{y,th}(\omega)$ are both white noises, the PLL frequency tracking scheme imposes the same filtering effect on them, with the only difference being the time constant $\tau_{th}$:  
\begin{equation}
H_{th}^{PLL}(j\omega) = \frac{(j\omega K_p + K_i)H_L(j\omega)}{-\omega^2+\frac{j\omega}{\tau_{th}}+(j\omega K_p + K_i)H_L(j\omega)}.
\label{eq:6}
\end{equation}
We then incorporate the PLL transfer functions [Eq.~(\ref{eq:4}-\ref{eq:6})] with $S_{y,add}^{int}(0)$ and $S_{y,th}^{int}(0)$ to obtain the noise after processing by the PLL frequency tracking: 
\begin{equation}
\begin{split}
S_{y,add}^{PLL}(\omega)=S_{y,mech}^{int}&(0)|H_{mech}^{PLL}(j\omega)|^2+\\&S_{y,det}(0)|H_{det}^{PLL}(j\omega)|^2,
\label{eq:7}
\end{split}
\end{equation}

\begin{equation}
S_{y,th}^{PLL}(\omega) = S_{y,th}^{int}(0)|H_{th}^{PLL}(j\omega)|^2.
\label{eq:8}
\end{equation}
Finally, the overall fractional frequency fluctuation $S_y(\omega)$ of a resonator under the PLL frequency tracking scheme is simply $S_{y}(\omega)=S_{y,add}^{PLL}(\omega)+S_{y,th}^{PLL}(\omega)$.

\begin{figure}[!htb]
\includegraphics[scale=0.85]{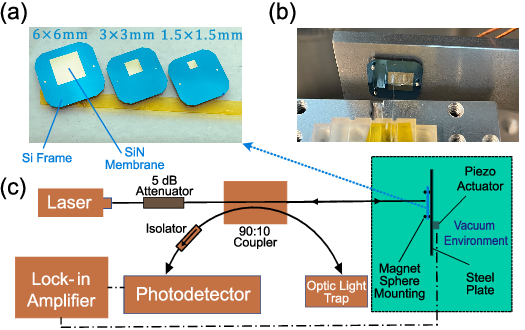}
\caption{\label{fig:two} \textbf{Schematics of the experimental setup} (a) Pictures of SiN drum resonators of three different sizes characterized in this work. (b) Picture of SiN drum resonator mounted on a steel plate using magnetic spheres inside our vacuum chamber. (c) Overall schematics of the experimental setup which includes the laser interferometer located outside of the high-vacuum chamber and the actuation and mounting method of SiN drum resonator inside of the chamber.}
\end{figure}

SiN drum resonators used in this work are fabricated using the process provided in \cite{snell_heat_2022}. During characterization, the SiN drum resonators are mounted magnetically onto a steel plate via three pairs of spherical magnets as shown in Fig.~\ref{fig:two}(b) and exited mechanically via a ceramic piezo actuator mounted on the other side of the steel plate [see Fig.~\ref{fig:two}(c)]. The magnet mounting method provides minimum contact area between the mounts and the chip, thus minimizing mechanical dissipation. We place the SiN drum resonators inside of a custom-built, high-vacuum ($8\times10^{-7}$ Torr typical operating pressure) chamber to minimize air damping (i.e., maintaining high-Q factor) and convective heat transfer. The chamber is left to thermally stabilize for two days (48 hours) after pumping down before we perform frequency fluctuation measurements. 

We detect the vibration signal of the SiN drum resonator, using a laser interferometer that consists of a $1550~\mathrm{nm}$ $\mathrm{Orion^{TM}}$ laser with built-in optical isolation, a 90:10 optical fiber coupler, 5 dB optical attenuator, optical isolator and a Thorlabs PDA20CS2 photodetector which is shown schematically in Fig.~\ref{fig:two}(c). The isolator at the location in Fig.~\ref{fig:two}(c) eliminates spurious optical cavities between the detector and the sample chip. The laser power output is set at $3.7~\mathrm{mW}$ which is then attenuated to $11.7~\mathrm{\mu W}$ prior to reaching the SiN drum resonator, via the 5 dB optical attenuator and the 90:10 optical coupler (i.e., 90\% power attenuation). Laser power attenuation is critical to minimize the effect of laser heating, as observed in a separate experiment in Supplemental Material \cite{sup}. The detected signal is sampled by a lock-in amplifier from Zurich Instrument Ltd. We use the phase-lock-loop (PLL) along with a proportional-integral controller provided by the lock-in amplifier to track the resonance frequency of the SiN drum resonators.

We quantify the frequency fluctuation of SiN drum resonators in this work using Allan deviation $\sigma_A$ \cite{allan_statistics_1966}, a metrology standard widely used to characterize the frequency fluctuation in nanoresonators. Based on the theoretically expected spectral density of frequency fluctuation $S_y(\omega)$, we can numerically compute the theoretical $\sigma_A$ via \cite{demir_understanding_2021}:
\begin{equation}
\sigma_A(\tau)=\frac{2}{\sqrt{\pi}\tau}\left[\int_{-\infty}^{\infty} \frac{[\sin(\frac{\omega\tau}{2})]^4}{\omega^2} S_y(\omega)\,d\omega \right]^{\frac{1}{2}},
\label{eq:9}
\end{equation}
where $\tau$ is the integration time. The asymptotic limit of $\sigma_A$ (i.e., excluding all intrinsic and PLL filters) for white noises can be computed analytically as $\sqrt{S_y(0)/\tau}$, such that the asymptote specifically for thermal fluctuation noise is $\sqrt{S_{y,th}^{PLL}(0)/\tau}$.

In order to observe additive phase noise and thermal fluctuation noise over a broad range of $\tau$, the demodulation time constant in our experiment is set to a high bandwidth ($\tau_{demod}=3.18\times10^{-5}$ s) to minimize signal filtering at the lock-in input. Likewise, the PLL bandwidth is set to $\approx5$ times faster than the thermal fluctuation bandwidth ($\tau_{PLL}\approx \tau_{th}/5$) to prevent filtering of thermal fluctuation noise $S_{y,th}^{int}$ by the PLL. The corresponding $K_p$ and $K_i$ values for achieving this bandwidth are determined using the relations given in \cite{demir_understanding_2021}. 

The geometric and material properties of the SiN drum resonators, and of the eigenmodes chosen for actuation, are decided to minimize additive phase noise related to thermal fluctuation noise according to Eq.~(\ref{eq:3}). The drum resonators large area [i.e., $L=1.5~\mathrm{mm}$, $3~\mathrm{mm}$, $6~\mathrm{mm}$ shown in Fig.~\ref{fig:two}(a)] maximizes the mass-to-thermal-conductance ratio $\frac{m_{eff}}{G}$ for the drum resonator thickness of 90 nm. The use of low-stress ($\sigma \approx 100$ MPa) SiN membrane maximizes the temperature sensitivity $\alpha$ via Eq.~(\ref{eq:3.1}). The values of $E$, $\alpha_T$ and $\nu$ for SiN are respectively 300 GPa, $2.2\times10^{-6}$ $\mathrm{K^{-1}}$ and 0.28, which result in $\alpha \approx$ $4.6\times10^{-3}$ $\mathrm{K^{-1}}$. We also pick high-order eigenmodes (high $f_r$) having high Q-factors. Specifically, the values of $f_r$ and Q-factors, from the smallest to largest SiN drum resonator are respectively 137.6, 90.2, 208.6 kHz and 0.86, 1.18, 1.24 million. 

\begin{figure}
\includegraphics[scale=0.8]{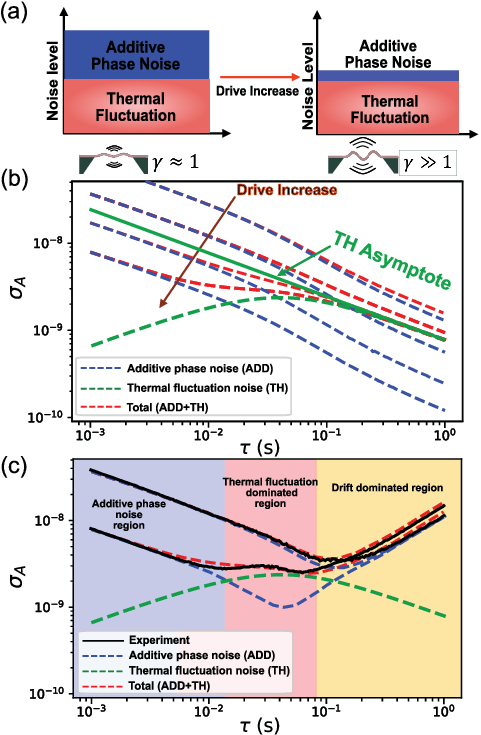}
\caption{\label{fig:three}(a) Schematic of changing noise profile via drive increase. (b) Theoretical Allan deviations $\sigma_A$ comparison between the model that solely considers additive phase noise (labelled "ADD"), the model that solely considers thermal fluctuation noise (labelled "TH") and the model that considers both additive phase noise and thermal fluctuation noise (labelled "ADD+TH"), for a $3\times3$ mm SiN drum resonator at four levels of $A_{rss}$. (c) Comparison of experimental Allan deviations $\sigma_A$ (black solid lines) with respect to all theoretical $\sigma_A$ at two levels of $A_{rss}$.}
\end{figure}

\begin{figure*}
\includegraphics[scale=1.17]{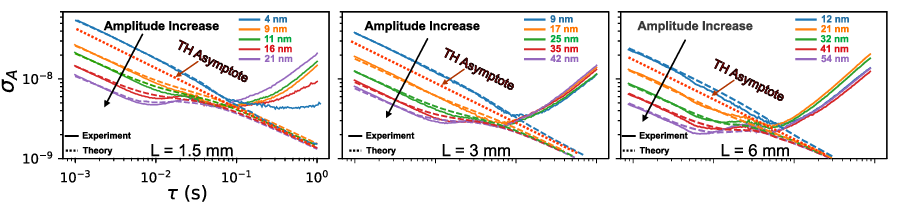}
\caption{\label{fig:four}Experimental and theoretical Allan deviations $\sigma_A$ for SiN drum resonators of three different sizes, at multiple vibration amplitudes $A_{rss}$.}
\end{figure*}

Considering these drum resonator parameters, we compare, in Fig.~\ref{fig:three}(b), the expected Allan deviation $\sigma_A$ plots for our model (labelled "ADD+TH") with that of recent model \cite{demir_understanding_2021} that solely includes additive phase noise (labelled "ADD"). In this case, we consider a low-stress $3\times3$ mm SiN drum resonator at different levels of actuation $A_{rss}$ and we set $\kappa_d = 0.012$ to account for typical detection noise in our experiment. We note that at small $\tau\lesssim0.005$ s, the two models overlap with each other, which indicates that additive phase noise is dominant. More specifically, $S_{y,det}$ dominates, since $S_{y,th}^{int}$ and $S_{y,mech}^{int}$ are attenuated by their respective intrinsic filter $H_{th}$ and $H_{mech}$ (with $\tau_{th}=\mathrm{0.09}$ s and $\tau_{mech}=\mathrm{4.15}$ s). At intermediate integration time $0.005\lesssim\tau\lesssim0.07$ s, and when $A_{rss}$ is sufficiently high, thermal fluctuation noise becomes non-negligible. In this case, theoretical $\sigma_A$ plots that include thermal fluctuation converge towards an $A_{rss}$-independent thermal fluctuation asymptote ($\sqrt{S_{y,th}^{PLL}(0)/\tau}$) as $\tau$ increases. On the contrary, considering only additive phase noise in Fig.~\ref{fig:three}(b) do not exhibit amplitude-independent-converging effect. 

This difference between the models is confirmed experimentally in Fig.~\ref{fig:three}(c), where we find that experimentally recorded $\sigma_A$ match closely with our model. Conversely, the "ADD" model fails at intermediate $\tau$ values (i.e., in the "thermal fluctuation dominated region") when $A_{rss}$ is sufficiently high for thermal fluctuation noise to dominate. Eventually, drift occurs and systematically dominates at $\tau\gtrsim0.07$ s. Note that values presented in Fig.~\ref{fig:three}(c) are a limited subset of the more complete data set presented in Fig.~\ref{fig:four}. The experimental conditions and model fitting parameters are discussed below for Fig.~\ref{fig:four} therefore also applies to Fig.~\ref{fig:three}(c).  

To further validate our theoretical model, we repeat Allan deviation $\sigma_A$ measurements for resonators of three different sizes (i.e., $L=1.5$ mm, 3 mm, 6 mm) and for several drive amplitudes $A_{rss}$ in Fig.~\ref{fig:four}. During experiments, we first excite the resonators at low $A_{rss}$, such that the experimental $\sigma_A$ is overall slightly above the thermal fluctuation asymptote---i.e., additive phase noise is marginally larger than thermal fluctuation noise. We then increase $A_{rss}$ progressively to reduce the additive phase noise. We find that all three drum resonators consistently converge towards the thermal asymptote at intermediate $\tau$ values, and when $A_{rss}$ is sufficiently large. The convergence is then rapidly shadowed by drift at large $\tau$.

Correspondence between the model and experiment in FIG.~\ref{fig:four} was obtained by using only fit parameters that are expected from our experimental uncertainties. Two fit parameters are used for all resonators, while a third one is needed only for the largest resonator. The vibration amplitude of the resonators $A_{rss}$ and the detection noise scaling factor $\kappa_d$ are fitted for all three resonators to account for misalignment uncertainly between the location of our optical fiber and the mode vibration anti-node. This misalignment is found to underestimates $A_{rss}$ by fitted factors of (1, 1.2, 2.5) for the three respective resonators sizes. Likewise, fitting $\kappa_d$ yields increase factors of (50\%, 30\%, 10\%) relative to the $\kappa_d$ values expected from our measurement noise floor ($\approx0.23$ $\mathrm{pm/\sqrt{Hz}}$) and from theoretically expected thermomechanical fluctuations. Finally, we find that for the largest resonator, the experimental thermal time constant $\tau_{th}$ must be reduced by a factor of 2 relative to the theoretically expected value. It is possible that the very high order mode used in this case lead to a higher $G$ (lower $\tau_{th}$) due to several mode anti-nodes being located close to the heat-dissipating silicon frame.

We note that our measurement in Fig.~\ref{fig:four} confirms important relations between the level of thermal fluctuation relative to the membrane dimensions and mechanical properties. We first observe the level of thermal fluctuation noise (i.e., the TH asymptote vertical position) scales inversely with the drum resonator side length (i.e., with $G$), as predicted by Eq.~(\ref{eq:2}) and Fig.~\ref{fig:five}. This as an important consequence in practice. While Eq.~(\ref{eq:2}) and Fig.~\ref{fig:five}(b) suggest that large drum resonators are always better (to maximize $\frac{m_{eff}}{G}$), this is not entirely true in practice. As our membrane size increases, it becomes increasingly difficult to identify the thermal asymptote before drifts occur at higher $\tau$. A trade-off therefore exists when designing sensors operating at the fundamental thermal fluctuation noise limit; large drum resonator should be used to maximize the $\frac{m_{eff}}{G}$ ratio up to the plateau observed in Fig.~\ref{fig:five}(b). However, thermal fluctuation noise limit will eventually be shadowed by drift if the resonator is too large ($L>>3$ mm).  

In conclusion, we present and experimentally validate a model for computing frequency stability of NMR considering the effect of both additive phase noise and thermal fluctuation noise in close-loop frequency tracking scheme. We demonstrated that by using SiN drum resonators of properly designed geometric and material properties, one can minimize additive phase noise below thermal fluctuation noise. We also identified which resonators inherent properties (e.g., thermal conduction, thermal coefficient of frequency) most affect thermal noise, in contrast with additive phase noise that can be minimized via other parameters (e.g., drive amplitude). Our work therefore provides fundamental guidance for building thermal sensors such as nanomechanical bolometers. We provide a path for those to reach the never attained fundamental detectivity limit \cite{Rogalski2003} of thermal radiation detection, which requires sensors limited only by fundamental thermal fluctuation noise \cite{snell_heat_2022}. 

The authors would like to acknowledge Prof. Silvan Schmid for useful discussion on laser-induced heating. This work was funded by Natural Sciences and Engineering Research Council of Canada (NSERC) Discovery Grants program.
\newpage


\nocite{*}
\bibliography{apssamp}% Produces the bibliography via BibTeX.
\newpage
\section*{Supplementary Information}
\addcontentsline{toc}{section}{Supplementary Information}
\renewcommand{\thefigure}{S\arabic{figure}}
\subsection{Finding the optimum laser power}
Optimizing the laser power is critical for our experiment: too high power may cause spurious heating that affects thermal fluctuation readings, while low power increases detection noise and can therefore prevent observation of fundamental thermal fluctuations. In this supplementary experiment, we empirically identify a suitable range of laser power for observing the dominant effect of thermal fluctuation noise in our SiN resonators. We replace the fixed optical attenuator in Fig.~\ref{fig:two}(c) by a variable optical attenuator (Thorlabs V1550A) such that the laser power incident on the SiN resonator can be varied. We then drive the SiN resonator to an amplitude (40 nm) suitable for observation of thermal fluctuations noise (see Fig.~\ref{fig:three}-\ref{fig:four}), and we record Allan deviations for different laser powers. As shown in Fig.~\ref{fig:s1}, in the thermal fluctuation region, the Allan deviation varies with laser power when the attenuation is the 0 -- 3.1 dB range, meaning that laser at this intensity heats up the membrane and adds to thermal fluctuation noise. In the optimal 3.1 -- 5 dB attenuation range, the characteristic bump of thermal fluctuations frequency noise is clearly noticeable and is now independent of laser power. We therefore chose a fixed 5 dB attenuation in this work. Above 5 dB attenuation, additive phase noise becomes excessive, such that the effect of thermal fluctuation cannot be observed anymore.
\begin{figure}[!htb]
    \setcounter{figure}{0}
    \centering
    \includegraphics[width=0.46\textwidth]{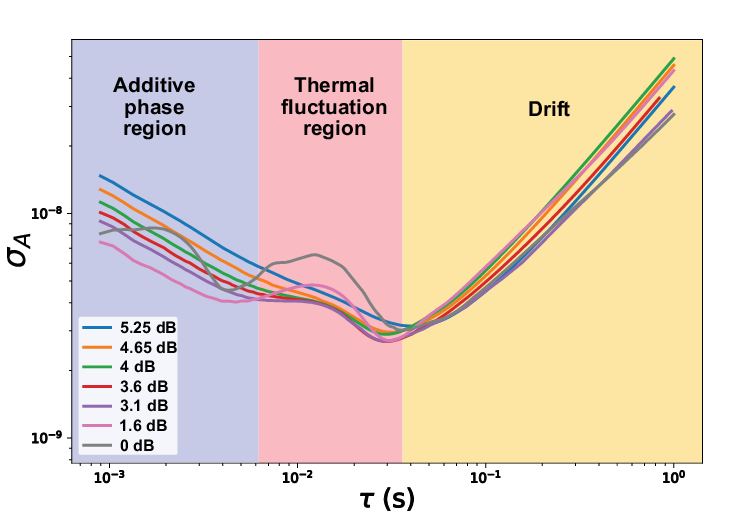}
    \caption{\label{fig:s1}Experimental Allan deviations $\sigma_A$ of a $3\times3$ mm SiN drum resonator at varying levels of laser power incident on the membrane. }
\end{figure}

\end{document}